\RequirePackage{fix-cm}
\documentclass[smallextended]{svjour3}       
\usepackage[italian,english]{babel}
\smartqed  
\usepackage{amsmath}
\usepackage{amssymb}
\usepackage{lipsum}
\usepackage{dsfont}
\usepackage{multirow}
\usepackage{afterpage}
\usepackage{amsfonts}
\usepackage{bm}
\usepackage{babel}
\usepackage{hyperref}
\usepackage{graphicx}
\usepackage{color}
\usepackage{graphicx}
\begin{document}

\title{Characterizing nonclassical correlations of tensorizing states in a bilocal scenario 
}


\author{S. Bhuvaneswari \and R. Muthuganesan   
\and  \\ R. Radha }


\institute{Centre for Nonlinear Science (CeNSc), PG \& Research Department of Physics, Government College for Women (Autonomous), Kumbakonam, Tamil Nadu, India\at
             \email{bhuvanajkm85@gmail.com}           \and 
Department of Physics, Faculty of Nuclear Sciences and Physical Engineering, \\ Czech Technical University in Prague,B\u rehov\'a 7, 115 19 Praha 1-Star\'e M\u{e}sto, \\ Czech Republic\at
              \email{rajendramuthu@gmail.com}           
           \and
         Centre for Nonlinear Science (CeNSc), PG \& Research Department of Physics, Government College for Women (Autonomous), Kumbakonam, Tamil Nadu, India\at
              \email{vittal.cnls@gmail.com}
}

\date{Received: date / Accepted: date}

\maketitle

\begin{abstract}
In the  present paper, we attempt to address the question of "can tensorizing states $(\rho\otimes\rho ~ \text{or}~\rho\otimes\rho')$" have quantum advantages?". To answer this question, we exploit the notion of measurement-induced nonlocality (MIN) and advocate a fidelity based nonbilocal measure to capture the nonlocal effects of tensorizing states due to locally invariant von Neumann projective measurements. We show that the properties of the fidelity based nonbilocal measure are retrieved from that of MIN. Analytically, we evaluate the nonbilocal measure for any arbitrary pure state. The upper bounds of the nonbilocal measure based on fidelity are also obtained in terms of eigenvalues of correlation matrix. As an illustration, we have computed the nonbilocality for some popular input states.
\end{abstract}

\keywords{Entanglement \and Quantum correlation  \and Dynamics \and Weak measurement \and Projective measurements \and Nonlocality}

\section{Introduction}
\label{intro}
The existence of nonlocal attributes in composite states continues to be a fundamental and unique feature of quantum systems which has no counterparts in the classical domain \cite{Nielsen2010}. In fact, it has fueled  the development of quantum technologies. A few notable nonlocal attributes are coherence \cite{Girolami2014,Baumgratz2014}, entanglement \cite{Einstein,scho,Bell} , steering and quantum correlations beyond entanglement \cite{Ollivier2001,Dakic,Luo2010pra,Luo2011}.  Ever since the identification of EPR paradox \cite{Einstein}, it is believed that the entanglement is the only manifestation of nonlocality of the quantum system. Recent studies have reinforced the fact that  entanglement  alone cannot capture the entire spectrum of the nonlocality and one will have to go beyond entanglement to get an estimate of the entire spectrum.  The nonlocality of pure states is completely characterized by the violation of Bell inequality and entanglement \cite{Werner}. On the other hand, the nonlocal aspects of mixed states are not understood and are still found to be mysterious entities in quantum information theory \cite{Werner,Almeida}. Despite the ongoing debate on the manifestation of nonlocality, the quantification of entanglement and quantum correlations beyond entanglement besides their characterization continue to capture the attention of researchers in the domain of quantum information processing. 

In a recent~ quantum entanglement swapping experiment with~ multi-measurements and multi-sources, it is shown that the independence of the sources can cause the nonlocal behaviors of probability distributions and is called nonbilocal correlations \cite{Branciard2010,Branciard2012}. These kinds of correlations are captured using nonlinear inequalities and one important class of these inequalities is the so-called binary-input-and-output bilocality inequality which is called the bilocality inequality \cite{Branciard2010,Branciard2012}. In recent times, some interesting progress has been made in this direction \cite{Fritz2012,Fritz2016,Wood2015,Henson,Chaves2015,Tavakoli2014,Tavakoli2016a,Tavakoli2016b,Rosset}. Gisin et al.  \cite{Gisin2017} have shown that a pair of entangled states can violate the bilocality inequality implying that tensorizing states may possess nonlocal correlations. There is a curiosity on understanding the nonlocal behavior of quantum systems when two bipartite states with vanishing correlations are combined and whether the tensorizing state  possesses nonlocal advantages or not.

The phenomena of combining two quantum systems show better quantum advantages than the individual counterparts. This is known as superactivation of nonlocality, symbolized as $0 + 0 > 0$ and it cannot occur in a purely classical world. The superactivation of nonlocality provides an answer for "can the state $\rho \otimes \rho$ be nonlocal if  $\rho$ is local". Recently,  the superactivation of quantum nonlocality in the sense of violating certain Bell inequalities with an entangled bound state is also demonstrated \cite{Palazuelos}. The same study is carried out in the context of tensor networks \cite{Cavalcanti2011,Cavalcanti2012}. Further, the superactivation was also considered for arbitrary entangled states by allowing local preprocessing on the tensor product of different quantum states ($\rho \otimes \rho'$)  symbolized as $1 + 0 > 1$ \cite{Masanes}. Since, the nonlocality inequalities are detect the presence of nonlocal aspects in the quantum systems and there is no analytical method to quantify this quantum correlation in bilocal scenario.

In order to characterize the  nonclassical correlation of bilocal states, we exploit the property of bipartite measurement-induced nonlocality (MIN) and define a nonbilocal measure using fidelity between the states.   The relation between the nonlocal and non-bilocal correlation measures is established and it is shown that non-bilocality is always greater than the nonlocal correlation. Further, the upper bounds of the nonbilocal correlation measure are also obtained for arbitrary mixed input states. To validate the properties of nonbilocal measures, we study the proposed quantity for  different input states.

The present paper is structured  as follows: To start with in Sec. \ref{MIN}, we  review the concept of measurement-induced nonlocality and definition of fidelity-based MIN. In Sec. \ref{Sec3}, we introduce fidelity-based nonbilocal measure and establish a relationship with the nonlocal measure. Sec. \ref{Sec4} quantifies the nonbilocality of arbitrary mixed states.  In Sec. \ref{Sec5}, we compute the proposed nonbilocal measure for some well known input states and compare with the fidelity-based MIN.  Finally, the conclusions are presented in Sec. \ref{Concl}. 

\section{Measurement-induced nonlocality}
\label{MIN}
Measurement-induced nonlocality (MIN) is manifested in the nonlocal effects due to locally invariant eigenprojective measurements and  is a faithful measure of bipartite quantum systems. It is originally defined as the maximal Hilbert-Schmidt distance between pre and post measurement states and is defined as \cite{Luo2011}
\begin{equation}
 N(\rho ) =~^{\text{max}}_{\Pi ^{a}}\| \rho - \Pi ^{a}(\rho )\| ^{2},  \label{HS-MIN}
\end{equation}
where the maximization is taken over the locally invariant projective measurements on subsystem $a$, $\Pi^{a}(\rho) = \sum _{k} (\Pi ^{a}_{k} \otimes   \mathds{1} ^{b}) \rho (\Pi ^{a}_{k} \otimes   \mathds{1}^{b} )$ with $\Pi ^{a}= \{\Pi ^{a}_{k}\}= \{|k\rangle \langle k|\}$ being the projective measurements on the subsystem $a$ which does not change the marginal state $\rho^{a}$ locally i.e., $\Pi ^{a}(\rho^{a})=\rho ^{a}$. Here $\|\mathcal{O} \|=\sqrt{\text{Tr}\mathcal{O}^{\dagger}\mathcal{O}} $  is the   Hilbert-Schmidt norm of operator $\mathcal{O} $. The MIN is in some sense dual to the  geometric version of quantum discord (GD) of the given state $\rho$ and is formulated as \cite{Dakic,Luo2010pra}
\begin{equation}
 D(\rho ) =~^{\text{min}}_{\Pi ^{a}}\| \rho - \Pi ^{a}(\rho )\| ^{2}. 
\end{equation}
If $\rho^{a}$ is nondegenerate, the optimization is not required and the above measures are equal. The Hilbert-Schmidt distance (HS) is an important quantity and its operational meaning is the informational distance between quantum states \cite{Lee2003}. Recently, it is shown that the HS norm is easy to compute and measurable experimentally using many-particle interference techniques \cite{Travnicek2019}.  Further, the Hilbert-Schmidt distance has been widely used in quantum mechanics, particularly in quantifying quantum resources, such as quantum entanglement \cite{Pandya2000}, quantum discord \cite{Dakic,Luo2010pra}, measurement-induced nonlocality \cite{Luo2011,Indrajith2021}, and asymmetry \cite{Yao2016}. The relations between the Hilbert-Schmidt distance and the trace distance have been studied \cite{Dodonov2000}. Nevertheless,  Hilbert-Schmidt distance is not a bonafide measure of quantum correlation which is shown by considering a simple map $\Gamma ^\sigma: X\rightarrow X \otimes\sigma$, i.e., the map adding a noisy ancillary state to the party $b$ \cite{Piani2012}. Under such an operation, we have
\begin{align}
\| X \| \rightarrow \| \Gamma ^\sigma X \|=\| X \| \sqrt{\text{Tr}\sigma^2}.
\end{align}
Due to the addition of local ancilla $\rho^c$, the MIN of resultant state is  
\begin{align}
 N(\rho^{a:bc} ) = N(\rho^{ab}) \text{Tr}(\rho ^{c})^2  \nonumber
\end{align}
implying that MIN differs arbitrarily due to local ancilla $c$  as long as $\rho^{c}$ is mixed. Defining the MIN in terms of any one of the contractive distance measures seems to be a natural way of rectifying the local ancilla problem. One such form of MIN  based on the fidelity is given by \cite{Hu2015}
\begin{equation}
 N_{F}(\rho) =1-~^{\text{min}}_{\Pi^{a}}~ F (\rho, \Pi ^{a}(\rho )).  \label{F-MIN}
\end{equation}
Here, the minimum is taken over the locally invariant projective measurements on subsystem $a$ and $F(\rho, \sigma)=\text{Tr}\sqrt{\rho^{1/2}\sigma\rho^{1/2}}$ is the fidelity between the states $\rho$ and $\sigma$ \cite{Jozsa1994}.  This measure has been explored in different contexts of quantum information processing such as cloning \cite{Gisin1997}, teleportation \cite{Zhang2007},  quantum chaos \cite{Gorin2006} and phase transition in physical systems \cite{Gu2010}. 

The above definition of fidelity involves the square root of density operator. Hence, the computation of fidelity in higher dimensions is quite intractable. To reduce the computational complexity, we follow another version of fidelity \cite{Wang}:
\begin{align}
  \mathcal{F}(\rho,\sigma)=\frac{\text{Tr}(\rho\sigma)^2}{\text{Tr}(\rho^2)\text{Tr}(\sigma^2)}.
\end{align}
This measure also possesses all the properties of fidelity introduced by Josza \cite{Jozsa1994} and is useful in defining MIN \cite{MuthuPLA}. Based on the fidelity, MIN is defined as \cite{MuthuPLA,MuthuPLA2}
\begin{equation}
 N_{\mathcal{F}}(\rho) =1-~^{\text{min}}_{\Pi^{a}} \mathcal{F} (\rho, \Pi ^{a}(\rho )).  \label{F-MIN2}
\end{equation}
It is worth  mentioning at this juncture that the fidelity-based MIN fixes the local ancilla problem of Hilbert-Schmidt MIN and satisfies all the necessary axioms of a bonafide measure of quantum correlations.

\section{Nonbilocality Measure}\label{Sec3}

In this section, we  introduce the fidelity-based nonbilocal measure. Let us consider a quantum state shared by four parties in the separable composite finite dimensional Hilbert space  $\mathcal{H}=\mathcal{H}_a\otimes\mathcal{H}_b\otimes\mathcal{H}_c\otimes\mathcal{H}_d$ with bipartition  $\rho_{ab}$ (shared between $a$ and $b$) and $\rho_{cd}$ (shared between $c$ and $d$). We define the nonbilocal measure in terms of fidelity as  
\begin{align}
N_{\mathcal{F}}(\rho_{ab}\otimes\rho_{cd})=1-~^{\text{min}}_{\Pi ^{bc}} \mathcal{F}(\rho_{ab}\otimes\rho_{cd}, \Pi^{bc}(\rho_{ab}\otimes\rho_{cd})),
\end{align}
where the optimization  is taken over the locally invariant eigenprojective measurements $\Pi^{bc}=\{ \Pi^{bc}_k\} $ which does not alter the marginal state $\rho^{bc}=\text{Tr}_{ad}(\rho_{ab}\otimes\rho_{cd})$   locally . $d_{\mathcal{F}}(\cdot, \cdot)$ quantifies the distance between the state and its post measurement state and  is given by $\Pi^{bc}(\sqrt{\rho_{ab}\otimes\rho_{cd}})=\sum_{k,l}(\mathds{1}^a \otimes \Pi^{bc}_{kl} \otimes\mathds{1}^d)\sqrt{\rho_{ab}\otimes\rho_{cd}}(\mathds{1}^a \otimes \Pi^{bc}_{kl} \otimes\mathds{1}^d)$ with $\mathds{1}^{a(d)}$ being a $2\times 2$ unit matrix acting on $a(d)$. Here, $\rho^b=\sum_i\lambda_i| i_b\rangle \langle i_b| $ and $\rho^c=\sum_j\lambda_j| j_c\rangle \langle j_c| $ are the marginal states of $\rho_{bc}$. If any one of the states is nondegenerate,  the measurement takes the form $\Pi^{bc}=\{\Pi^{b}\otimes\Pi^{c} \} $. 

Using the orthogonality of projectors and cyclic property of trace of matrices, we can show that $\text{Tr}(\Pi^{bc}(\rho_{ab}\otimes\rho_{cd}))^2=\text{Tr}(\rho_{ab}\otimes\rho_{cd} \Pi^{bc}(\rho_{ab}\otimes\rho_{cd}))$. Hence, the definition of nonbilocal measure can be recast as 
\begin{align}
N_{\mathcal{F}}(\rho_{ab}\otimes\rho_{cd})=1-~^{\text{min}}_{\Pi ^{bc}}\frac{\text{Tr}(\rho_{ab}\otimes\rho_{cd} \Pi^{bc}(\rho_{ab}\otimes\rho_{cd}))}{\text{Tr}(\rho_{ab}\otimes\rho_{cd})^2}. 
\label{nonbilocal}
\end{align}

 The above measure quantifies the quantum correlation of any bilocal state from the perspectives of eigenprojective measurements.  It is worth mentioning to MIN is a useful resource for bipartite quantum communication. In view of this,  the nonbilocal measure is also helpful for multipartite communication protocol. In general, the bilocal states are useful in typical entanglement swapping and the fidelity based nonbilocal measure involves in swpping of quantum resources. The nonbilocal measure given by Eq. (\ref{nonbilocal}) based on fidelity enjoys a variety of nice properties:
\begin{enumerate}
\item[(i)] $N_{\mathcal{F}}(\rho_{ab}\otimes\rho_{cd})\geq 0$ and the equality holds  for any product input states defined by $\rho_{ab}=\rho^a\otimes\rho^b$ and $\rho_{cd}=\rho^c\otimes\rho^d$. Also, the nonbilocal measure vanishes for classical-quantum state $\rho_{ab}=\sum_i\rho^a_i\otimes p_i| i_b\rangle \langle i_b|$ and $\rho_{cd}=\sum_j q_j| j_c\rangle \langle j_c|\otimes \rho^d_j$.

\item[(ii)]   $N_{\mathcal{F}}(\rho_{ab}\otimes\rho_{cd})$  is locally unitary invariant in the sense that 
\begin{align}
N_{\mathcal{F}}((U_{ab}\otimes U_{cd})\rho_{ab}\otimes\rho_{cd}(U_{ab}\otimes U_{cd})^{\dagger})=N_{\mathcal{F}}(\rho_{ab}\otimes\rho_{cd}),
\end{align}
where $U_{ab}=U_a\otimes U_b$ and $U_{cd}=U_c\otimes U_d$ are the local unitary operators.
\item[(iii)] If $N_{\mathcal{F}}(\rho_{ab}\otimes\rho_{cd})$  is positive, then, it implies that at least any one of the input states is entangled. 

\item[(iv)] If $\rho^b$ and $\rho^c$ are both nondegerate, then 
\begin{align}
N_{\mathcal{F}}(\rho_{ab}\otimes\rho_{cd})=1-~ \mathcal{F}(\rho_{ab}\otimes\rho_{cd}, \Pi^{bc}(\rho_{ab}\otimes\rho_{cd})). \nonumber
\end{align}
\item[(v)] For any pure input state, the fidelity-based nonbilocal measure is given by
\begin{align}
N_{\mathcal{F}}(|\Psi_{ab}\rangle\otimes|\Psi_{cd}\rangle)=1-\sum_{i,j}s_i^4r_j^4 \nonumber
\end{align}
where $s_i$ and $r_j$ are the Schmidt coefficients of $|\Psi_{ab}\rangle$ and $|\Psi_{cd}\rangle$ respectively. 
\item[(vi)] For any arbitrary pure state, the fidelity-based MIN and nonbilocal measures are related as 
\begin{align}
N_{\mathcal{F}}(\rho_{ba}\otimes\rho_{ab}) \geq N_{\mathcal{F}}^{\text{MIN}}(\rho). \nonumber
\end{align}
\item[(vii)] Although $N_{\mathcal{F}}(\rho_{ab})=N_{\mathcal{F}}(\rho_{cd})=0$, $N_{\mathcal{F}}(\rho_{ab}\otimes\rho_{cd})> 0$.
\end{enumerate}
 
The properties (i)-(iv) can be easily  proved. Hence, we concentrate on  the detailed proof of the remaining properties.
 
To validate property (v), we employ the Schmidt decomposition of pure input states.  Let $|\Psi_{ab}\rangle$ and $|\Psi_{cd}\rangle$ be the pure input states with the following Schmidt decomposition $|\Psi_{ab}\rangle=\sum_i\sqrt{s_i}|i_a  i_b \rangle$  and $|\Psi_{cd}\rangle =\sum_j\sqrt{r_j}|j_c j_d\rangle$  and $s_i$ and $r_j$ are the respective Schmidt coefficients of input states.  Further, $|i_{a(b)} \rangle$ and $|j_{c(d)} \rangle$ are the orthonormal bases of the subsystems $a(b)$ and $c(d)$ respectively

We note that 
\begin{equation}
\begin{aligned}
\rho_{ab}\otimes \rho_{cd}&=|\Psi_{ab}\rangle\langle\Psi_{ab}|\otimes|\Psi_{cd}\rangle\langle\Psi_{cd}|\\
&=\sum_{ii^{'}jj^{'}}s_is_{i^{'}}r_jr_{j^{'}}|i_a\rangle\langle i^{'}_a|\otimes |i_b\rangle\langle i^{'}_b|\otimes |j_c\rangle\langle j^{'}_c|\otimes |j_d\rangle\langle j^{'}_d|.
\end{aligned}
\end{equation}
For pure input states $\text{Tr}(\rho_{ab}\otimes\rho_{cd})^2=1$, the nonbilocal measure takes the form
\begin{align}
N_{\mathcal{F}}(|\Psi_{ab}\rangle\otimes|\Psi_{cd}\rangle)=1-~^{\text{min}}_{\Pi ^{bc}}\text{Tr}(\rho_{ab}\otimes\rho_{cd} \Pi^{bc}(\rho_{ab}\otimes\rho_{cd})).
\end{align}
Further, we compute the marginal state as 
\begin{equation}
\begin{aligned}
\rho^{bc}&=\text{Tr}_{ad}(|\Psi_{ab}\rangle\langle\Psi_{ab}|\otimes|\Psi_{cd}\rangle\langle\Psi_{cd}|)=\sum_{ij}s_i^2r_j^2|i_bj_c\rangle\langle i_bj_c|.
\label{marginal}
\end{aligned}
\end{equation}
The von Neumann projective measurement is expressed as 
\begin{align}
\Pi^{bc}=\{\Pi^{bc}_{hk}\equiv U|h_{b}k_{c}\rangle\langle h_{b}k_{c}|U^{\dag}\}
\label{measure}
\end{align}
and the marginal state can be expressed as a spectral decomposition of $\rho^{bc}$ as
$$\rho^{bc}=\sum_{hk}\langle h_{b}k_{c}|U^{\dag}\rho^{bc}U|h_{b}k_{c}\rangle U|h_{b}k_{c}\rangle\langle h_{b}k_{c}|U^{\dag}$$.
We wish to point out that  $\{U|h_{b}k_{c}\rangle\}$ is an orthonormal base with the eigenvalue $\langle h_{b}k_{c}|U^{\dag}\rho^{bc}U|h_{b}k_{c}\rangle$. 

The post-measurement state $\Pi^{bc}(\rho_{ab}\otimes \rho_{cd})$ can be computed as
\begin{equation*}
\begin{aligned}
&\Pi^{bc}(\sqrt{\rho_{ab}\otimes \rho_{cd}})=\Pi^{bc}(\rho_{ab}\otimes \rho_{cd})\\
=&\sum_{hk}(\mathds{1}^{a}\otimes\Pi_{hk}^{bc}\otimes \mathds{1}^{d})(|\Psi_{ab}\rangle\langle\Psi_{ab}|\otimes|\Psi_{cd}\rangle\langle\Psi_{cd}|)(\mathds{1}^{a}\otimes\Pi_{hk}^{BC}\otimes \mathds{1}^{d})\\
=&\sum_{hk}(\mathds{1}^{a}\otimes\Pi_{hk}^{bc}\otimes \mathds{1}^{d})(\sum_{ii^{'}jj^{'}}s_is_{i^{'}}r_jr_{j^{'}}|i_a\rangle\langle i^{'}_a|\otimes |i_b\rangle\langle i^{'}_b|\otimes |j_c\rangle\langle j^{'}_c|\otimes|j_d\rangle\langle j^{'}_d|)\\\notag 
\;\;\;\;\;\;\;\;&(\mathds{1}^{a}\otimes\Pi_{hk}^{bc}\otimes \mathds{1}^{d})\\
=&\sum_{hk}\sum_{ii^{'}jj^{'}}s_is_{i^{'}}r_jr_{j^{'}}|i_a\rangle\langle i^{'}_a|\otimes \Pi^{bc}_{hk}|i_b j_{c}\rangle\langle i^{'}_b j^{'}_c|\Pi^{bc}_{hk}\otimes |j_d\rangle\langle j^{'}_d|\\
=&\sum_{hk}\sum_{ii^{'}jj^{'}}s_is_{i^{'}}r_jr_{j^{'}}|i_a\rangle\langle i^{'}_a|\otimes U|h_{b}k_{c}\rangle\langle h_{b}k_{c}|U^{\dag}|i_b j_{c}\rangle\langle i^{'}_b j^{'}_c|U|h_{b}k_{c}\rangle\\ \notag
&\langle h_{b}k_{c}|U^{\dag}\otimes |j_d\rangle\langle j^{'}_d|.
\end{aligned}
\end{equation*}

Consequently, we have
\begin{equation*}
\begin{aligned}
&\rho_{ab}\otimes \rho_{cd}\Pi^{bc}(\rho_{ab}\otimes \rho_{cd})\\
=&(\sum_{ii^{'}jj^{'}}s_is_{i^{'}}r_jr_{j^{'}}|i_a\rangle\langle i^{'}_a|\otimes |i_b\rangle\langle i^{'}_b|\otimes |j_c\rangle\langle j^{'}_c|\otimes |j_d\rangle\langle j^{'}_d|)(\sum_{hk}\sum_{uu^{'}vv^{'}}s_us_{u^{'}}r_vr_{v^{'}}\\
&\langle h_{b}k_{c}|U^{\dag}|u_b v_{c}\rangle\langle u^{'}_b v^{'}_c|U|h_{b}k_{c}\rangle|u_a\rangle\langle u^{'}_a|\otimes U|h_{b}k_{c}\rangle\langle h_{b}f_{c}|U^{\dag}\otimes |v_d\rangle\langle v^{'}_d|)\\
=&\sum_{ii^{'}jj^{'}}\sum_{hk}\sum_{uu^{'}vv^{'}}s_is_{i^{'}}r_jr_{j^{'}}s_us_{u^{'}}r_vr_{v^{'}}\langle e_{B}f_{C}|U^{\dag}|u_B v_{C}\rangle\langle u^{'}_b v^{'}_c|U|h_{b}k_{f}\rangle|i_a\rangle\\
&\langle i^{'}_a|u_a\rangle\langle u^{'}_a|\otimes|i_b j_c\rangle\langle i^{'}_b j^{'}_c|U|h_{b}k_{c}\rangle\langle h_{b}k_{c}|U^{\dag}\otimes |j_d\rangle\langle j^{'}_d|v_d\rangle\langle v^{'}_d|.
\end{aligned}
\end{equation*}
Then, the fidelity between the pre- and post-measurement states is computed as 
\begin{equation*}
\begin{aligned}
&\mathcal{F}(\rho_{ab}\otimes \rho_{cd},\Pi^{bc}(\rho_{ab}\otimes \rho_{cd}))=\text{Tr}\rho_{ab}\otimes \rho_{cd}\Pi^{bc}(\rho_{ab}\otimes \rho_{cd})\\
=&\sum_{iujvhk}s_i^2s_{u}^2r_j^2r_{v}^2\langle h_{b}k_{c}|U^{\dag}|u_b v_{c}\rangle\langle i_b j_c|U|h_{b}k_{c}\rangle\langle u_b v_c|U|h_{b}k_{c}\rangle\langle h_{b}k_{c}|U^{\dag}|i_b j_c\rangle\\
=&\sum_{hk}(\langle h_{b}k_{c}|U^{\dag}\rho_{bc} U|h_{b}k_{c}\rangle)^{2}.
\end{aligned}
\end{equation*}
The nonbilocal measure for any pure state is given by 
\begin{equation*}
\begin{aligned}
N_{\mathcal{F}}(|\Psi_{ab}\rangle\otimes|\Psi_{cd}\rangle)=& \max_{\Pi^{bc}} d_{\mathcal{F}}(\rho_{ab}\otimes\rho_{cd},\Pi^{bc}(\rho_{ab}\otimes \rho_{cd})) \\
=& 1-\min_{\Pi^{bc}} \mathcal{F}(\rho_{ab}\otimes \rho_{cd},\Pi^{bc}(\rho_{ab}\otimes \rho_{cd}))\\
=&1-\min_{\Pi^{bc}}\sum_{hk}(\langle h_{b}k_{c}|U^{\dag}\rho^{bc} U|h_{b}k_{c}\rangle)^{2},
\end{aligned}
\end{equation*}
where the optimization is taken over all locally invariant measurements given in Eq. (\ref{measure}) leaving the marginal state $\rho^{bc}$ invariant. Thus, we obtain the property (v)
\begin{align}
N_{\mathcal{F}}(|\Psi_{ab}\rangle\otimes|\Psi_{cd}\rangle)=1-\sum_{i,j}s_i^4 r_j^4.
\end{align}
Hence,  proved.

%

 To establish the property (vi), we first recall the definition of fidelity based nonbilocal correlation measure as 
\begin{align}
N_{\mathcal{F}}(\rho_{ba}\otimes\rho_{ab}) =&~^{\text{~max}}_{\Pi ^{aa}} ~d_{\mathcal{F}}(\rho_{ba}\otimes\rho_{ab}, \Pi^{aa}(\rho_{ba}\otimes\rho_{ab})), \nonumber \\
=&1-~^{\text{min}}_{\Pi ^{aa}} \mathcal{F}(\rho_{ba}\otimes\rho_{ab}, \Pi^{aa}(\rho_{ba}\otimes\rho_{ab})), \nonumber \\
=&1-~^{\text{min}}_{\Pi ^{bc}}\frac{\text{Tr}(\rho_{ab}\otimes\rho_{cd} \Pi^{bc}(\rho_{ab}\otimes\rho_{cd}))}{\text{Tr}(\rho_{ab}\otimes\rho_{cd})^2}. \nonumber
\end{align} 
For pure input states, $\text{Tr}(\rho_{ab}\otimes\rho_{cd})^2=1$ and the nonbilocal correlation measure is given by
\begin{align}
N_{\mathcal{F}}(\rho_{ba}\otimes\rho_{ab}) =&1-~^{\text{min}}_{\Pi ^{aa}}\text{Tr}(\rho_{ab}\otimes\rho_{cd} \Pi^{bc}(\rho_{ab}\otimes\rho_{cd})) \nonumber \\
\geq& 1- ~^{\text{~min}}_{\Pi ^{a}} ~\text{Tr}(\rho_{ba}\otimes\rho_{ab} (\Pi^{a}\otimes\Pi^{a})(\rho_{ba}\otimes\rho_{ab})). \nonumber \\
=&1-~^{\text{~min}}_{\Pi ^{a}} \text{Tr}(\rho_{ab} \Pi^a(\rho_{ab}) )^2 \nonumber \\
\geq& 1-~^{\text{~min}}_{\Pi ^{a}} \text{Tr}(\rho_{ab} \Pi^a(\rho_{ab}) ) \nonumber \\
=&N_{\mathcal{F}}^{MIN}(\rho), \nonumber
\end{align} 
where the first inequality follows from the fact that $\Pi^a\otimes \Pi^a$ is not necessarily optimal and the second inequality is due to the square of the fidelity between pre- and post- measurement states and is either equal to or  less than unity. Hence, the theorem is proved. The above relation provides a closer connection between the nonbilocal and nonlocal measures  implying that the nonbilocal measure is always greater than MIN.  

\section{Nonbilocal correlation for Mixed states}
\label{Sec4}
To compute the fidelity-based nonbilocal measure for any arbitrary mixed input state, we first define some basic notation in the operator Hilbert space. Let $\mathcal{L}(\mathcal{H}_{\alpha})$ be the Hilbert space of linear operators on $\mathcal{H}_{\alpha} (\alpha=a,b,c,d)$ with the inner product $\langle X|Y\rangle=\text{Tr}X^{\dagger}Y $. An arbitrary  $m\times n$ dimensional bipartite state can be written as
\begin{align}
\rho_{ab}=\sum_{i,j} \lambda^{ab}_{ij} X_i\otimes Y_j, \nonumber
\end{align}
where $\{X_i : i=0,1, \cdots, m^2-1 \}$ and  $\{Y_j : j=0,1, \cdots, n^2-1 \}$ are the orthonormal operator bases of the subsystems $a$ and $b$ respectively satisfying the relation $\text{Tr}X_kX_l=\delta_{kl}$ and $\lambda^{ab}_{ij}=\text{Tr}(\rho_{ab}~X_i\otimes Y_j)$ are real elements of matrix $\Lambda_{ab}$. Similarly, one can define the orthonormal operator bases  as $\{P_k : k=0,1, \cdots, u^2-1 \} ~\text{and~} \{Q_l : l=0,1, \cdots, v^2-1 \}$ for another input state $\rho_{cd}$ with $u$ and $v$ being the dimensions of the marginal systems $c$ and $d$ respectively.  Then, the state $\rho_{cd}$ is defined as 
\begin{align}
\rho_{cd}=\sum_{k,l} \lambda^{cd}_{kl} P_k\otimes Q_l,  \nonumber
\end{align}
where $\lambda^{cd}_{kl}=\text{Tr}(\rho_{cd}~P_k\otimes Q_l)$ are the matrix elements of matrix $\Lambda_{cd}$. Then, the bilocal state is written as 
\begin{align}
\rho_{ab}\otimes \rho_{cd}=\sum_{i,j} \sum_{k,l} \lambda^{ab}_{ij}\lambda^{cd}_{kl}  X_i\otimes Y_j\otimes P_k\otimes Q_l.
\label{bilocal}
\end{align}

\textit{{\bf Theorem 1}: For any arbitrary bilocal input states represented in Eq. (\ref{bilocal}), the upper bound of nonbilocal measure is given by
\begin{align}
N_{\mathcal{F}}(\rho_{ab}\otimes\rho_{cd})\leq 1-\frac{1}{\| \Lambda\|^2 }\sum_{s=1}^{nu}\mu_{s},
\end{align}
where $\mu_{s}$ are the eigenvalues of the matrix $\Lambda_{ab,cd}\Lambda^t_{ab,cd}$ arranged in increasing order and $\Lambda^t_{ab,cd}$ denotes the transpose of the  matrix $\Lambda_{ab,cd}$. 
}

If the measurement operators are given by $\Pi^{bc}=\{ \mathds{1}^a\otimes\Pi^{bc}_h\otimes\mathds{1}^d\} $, the post-measurement state becomes
\begin{align}
\Pi^{bc}(\rho_{ab}\otimes \rho_{cd})=&\sum_h \sum_{ijkl}  \lambda^{ab}_{ij}\lambda^{cd}_{kl} X_i\otimes\Pi^{bc}_h(Y_j\otimes P_k)\Pi^{bc}_h\otimes Q_l \nonumber \\
=& \sum_h \sum_{ijj'kk'l}  \lambda^{ab}_{ij}\lambda^{cd}_{kl} \gamma_{hjk} \gamma_{hj'k'} X_i\otimes Y_{j'}\otimes P_{k'}\otimes Q_l,
\end{align}
where $\gamma_{hjk}=\text{Tr}\Pi^{bc}_h(Y_j\otimes P_k)$ are the elements of matrix $\Gamma$. Next, we compute the fidelity between pre- and post-measurement states and is given by
\begin{align}
\mathcal{F}(\rho_{ab}\otimes \rho_{cd},\Pi^{bc}(\rho_{ab}\otimes \rho_{cd}))=&\sum_h \sum_{ijj'kk'l} ~ \lambda^{ab}_{ij}\lambda^{cd}_{kl} \gamma_{hjk} \gamma_{hj'k'}\lambda^{ab}_{ij'}\lambda^{cd}_{k'l} \nonumber \\
=&\frac{1}{\| \Lambda\|^2 } \Gamma \Lambda_{ab,cd}\Lambda^t_{ab,cd} \Gamma^t,
\end{align}
where $\Gamma$ is an $nu\times n^2 u^2$   dimensional matrix. Then, 
\begin{align}
N_{\mathcal{F}}(\rho_{ab}\otimes\rho_{cd})=1-~^{\text{min}}_{~\Gamma}~\frac{1}{\| \Lambda\|^2 }\Gamma \Lambda_{ab,cd}\Lambda^t_{ab,cd} \Gamma^t\leq 1-\frac{1}{\| \Lambda\|^2 }\sum_{s=1}^{nk}\mu_{s}, \nonumber
\end{align}
where $\mu_{s}$ are the eigenvalues of the matrix $\Lambda_{ab,cd}\Lambda^t_{ab,cd}$ listed in increasing order. Hence the theorem is proved.

\textit{{\bf Theorem 2}: If the marginal state $\rho^b$ is nondegenerate,  the nonbilocal measure $N_{\mathcal{F}}(\rho_{ab}\otimes\rho_{cd})$  due to the measurement $\Pi^{bc}$ has the upper bound as 
\begin{align}
N_{\mathcal{F}}(\rho_{ab}\otimes\rho_{cd})\leq 1- \frac{1}{\| \Lambda_{cd}\|^2}{\mathcal{F}}(\rho_{ab}, \Pi^b(\rho_{ab})) \times (\sum_{\tau=1}^{u}\mu_{\tau}),
\end{align}
where $\mu_{\tau}$ are the eigenvalues of matrix $\Lambda_{cd}\Lambda^t_{cd}$ arranged in increasing order and $\mathcal{F}(\rho_{ab}, \Pi^b(\rho_{ab}))$ is the fidelity between the state $\rho_{ab}$ and post-measurement state $\Pi^{b}(\sqrt{\rho_{ab}})$. 
}

To prove the theorem, it is worth reiterating that if the marginal state is nondegenerate, the optimization is not required, Assuming that the state $\rho^b$ is nondegenerate and the optimization of the measure given by Eq. (\ref{nonbilocal}) is taken over $\Pi^c$ alone, the measurement operator is defined as $\Pi^b\otimes\Pi^c=\{ \Pi^b_j\otimes\Pi^c_k\}=\{| j_b\rangle \langle j_b| \otimes \Pi^c_k\} $. The nonbilocality measure based on the fidelity becomes
\begin{align}
N_{\mathcal{F}}(\rho_{ab}\otimes\rho_{cd})=&1-~^{\text{min}}_{\Pi ^{bc}} ~\mathcal{F}(\rho_{ab}\otimes\rho_{cd},\Pi ^{bc}(\rho_{ab}\otimes\rho_{cd})) \nonumber \\
=& 1-~^{\text{min}}_{\Pi ^{bc}} ~\frac{\text{Tr}\rho_{ab}\otimes\rho_{cd}\cdot\Pi ^{bc}(\rho_{ab}\otimes\rho_{cd})}{\text{Tr}(\rho_{ab}\otimes\rho_{cd})^2} \nonumber \\
=& 1-~^{\text{min}}_{\Pi ^{c}}~\frac{\text{Tr}\rho_{ab}\otimes\rho_{cd}\cdot(\Pi^{b}\otimes\Pi ^{c})(\rho_{ab}\otimes\rho_{cd})}{\text{Tr}\rho_{ab}^2\cdot\text{Tr}\rho_{cd}^2} \nonumber \\
=& 1-\frac{1}{\| \Lambda_{cd}\|^2}{\mathcal{F}}(\rho_{ab}, \Pi^b(\rho_{ab})) ~^{\text{min}}_{\Pi ^{c}}~\text{Tr}~\rho_{cd}\Pi ^{c}(\rho_{cd}).
\end{align}
where $\mathcal{F}(\rho_{ab}, \Pi^b(\rho_{ab}))$ is the fidelity between the state $\rho_{ab}$ and post-measured state $\Pi^b(\rho_{ab})$.  Following the optimization procedure given in \cite{MuthuPLA},  we write the second term of the above equation as 
\begin{align}
~^{\text{min}}_{~\Pi ^{c}}~\text{Tr}~\rho_{cd}\Pi ^{c}(\rho_{cd})=~^{\text{min}}_{~C}~\text{Tr}~C\Lambda_{cd}\Lambda_{cd}^tC^t.
\end{align}
The quantity $\text{Tr}\rho_{ab}\Pi^{b}(\rho_{ab})$ is the fidelity between the state $\rho_{ab}$ and post-measurement state $\Pi^{b}(\rho_{ab})$. Then, the fidelity-based nonbilocal measure is given by
\begin{align}
N_{\mathcal{F}}(\rho_{ab}\otimes\rho_{cd})=&1-\mathcal{F}(\rho_{ab},\Pi^{b}(\rho_{ab}))~^{\text{min}}_{~C}~\text{Tr}~C\Lambda_{cd}\Lambda_{cd}^tC^t \nonumber \\
\leq & 1- 1-\frac{1}{\| \Lambda_{cd}\|^2 }\mathcal{F}(\rho_{ab}, \Pi^b(\rho_{ab})) \times \sum_{\tau=1}^{u}\mu_{\tau},
\end{align}
where $\mu_{\tau}$ are the eigenvalues of matrix $\Lambda_{cd}\Lambda^t_{cd}$ arranged in increasing order.

If the marginal states $\rho^b$ and $\rho^c$ are nondegenerate and the dimension of  $\rho^c$ is 2 $(u=2)$, then, the closed formula of nonbilocal measure is expressed as 
\begin{align}
N_{\mathcal{F}}(\rho_{ab}\otimes\rho_{cd})\leq 1- {\mathcal{F}}(\rho_{ab}, \Pi^b(\rho_{ab})) \times \frac{(\mu_{1}+\mu_{1})}{\| \Lambda_{cd}\|^2}.
\end{align}

\section{Illustrations}
\label{Sec5}
In this section, we compute the fidelity-based measurement-induced nonbilocality for some well known input states. 

\textit{Example 1:} Let $|\Psi_{ab}\rangle=|00\rangle $ and $|\Psi_{cd}\rangle=(|00\rangle+ |11\rangle)/\sqrt{2}$ be the two input states. 
According to property (v), the nonbilocal measure is 
\begin{align}
N_{\mathcal{F}}(|\Psi_{ab}\rangle\otimes|\Psi_{cd}\rangle)=1-\sum_{i,j}s_i^4r_j^4. 
\end{align}
The Schmidt coefficients for $|\Psi_{ab}\rangle$ are 0 and 1. Similarly, $|\Psi_{cd}\rangle$ has the Schmidt coefficients $1/\sqrt{2}$ and $1/\sqrt{2}$. Then, $N_{\mathcal{F}}(|\Psi_{ab}\rangle\otimes|\Psi_{cd}\rangle)=0.5$. The above example validates the property (iii) of $N_{\mathcal{F}}(\rho_{ab}\otimes\rho_{cd})$.

\textit{Example 2:} The input state is $|\Psi_{ab}\rangle\otimes|\Psi_{cd}\rangle=(|00\rangle+|11\rangle)/\sqrt{2}\otimes(|00\rangle+|11\rangle)/\sqrt{2}$. Then, 
\begin{align}
N_{\mathcal{F}}(|\Psi_{ab}\rangle\otimes|\Psi_{cd}\rangle)=1-4 \times \frac{1}{4} \times \frac{1}{4} =\frac{3}{4}.
\end{align}
\textit{Example 3:} Next, we consider the isotropic state given by
\begin{align}
\rho=\frac{1-x}{m^2-1}\mathds{1}+\frac{m^2x-1}{m^2-1}| \Psi \rangle \langle \Psi|; ~~~~x\in [0,1].
\end{align}
where $\Psi=\frac{1}{\sqrt{m}}\sum_i | ii \rangle$ is a maximally entangled state and $m$ is the dimension of the state. 

Here, we compare the F-MIN and nonbilocal measure. The F-MIN is zero when $x=1/m^2$ and the corresponding state is a maximally mixed state. We have computed the fidelity based nonlocal and nonbilocal measures and plotted them as a function of state parameter $x$ in Fig. (1). From Fig. (1), we observe that the nonbilocal measure is also zero at $x=1/m^2$. Further, we notice that nonbilocal measure is always greater than nonlocal measure (F--MIN).

\begin{figure*}[!ht]
\centering\includegraphics[width=0.5\linewidth]{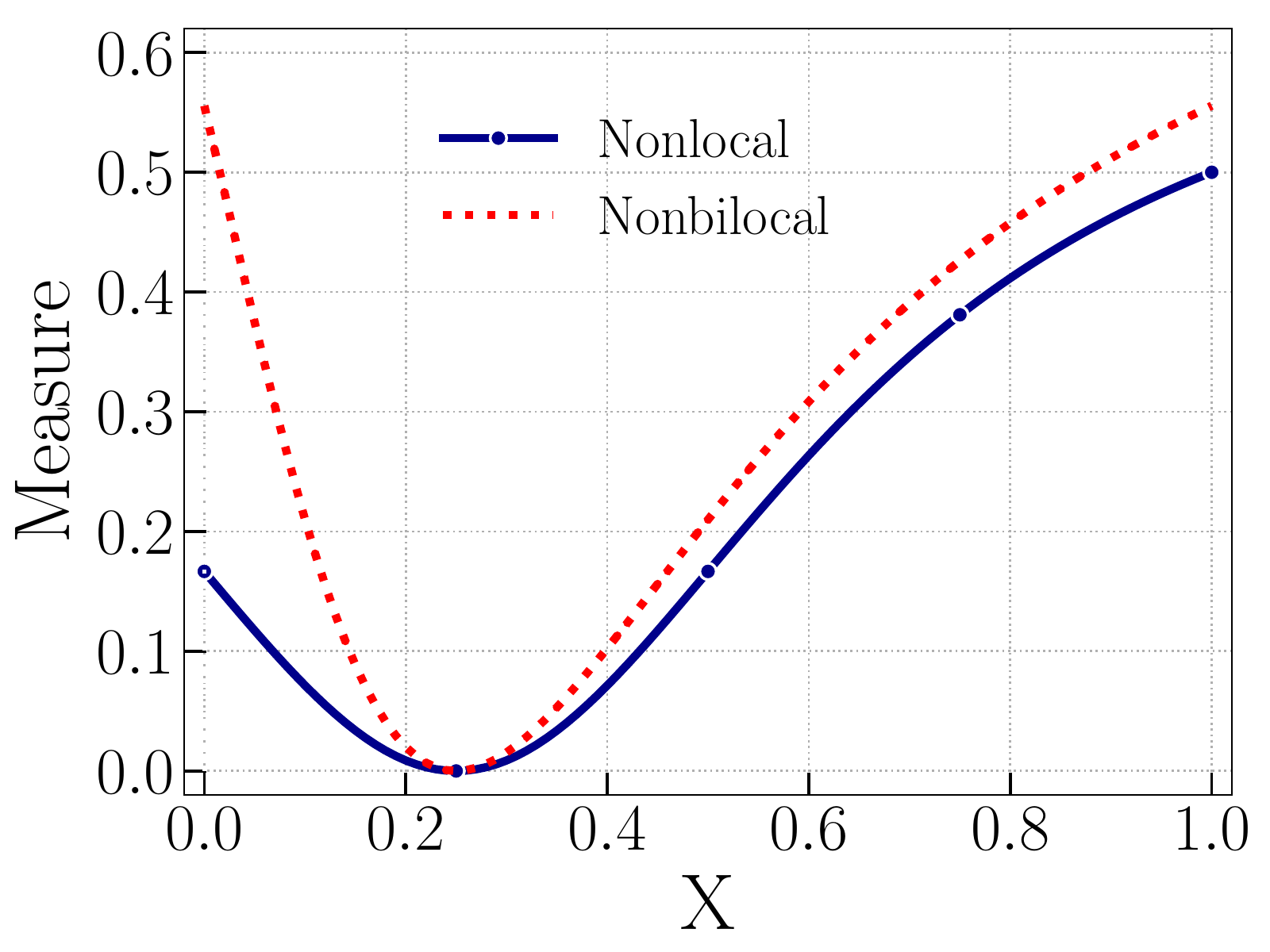}
\caption{(color online) The fidelity based nonlocal and nonbilocal measure for a $2\times 2$ isotropic state. }
\label{fig1}
\end{figure*}

\textit{Example 4:} Next, we study the nonbilocality of  the $m\times m$ dimensional Werner state  and  is given by
\begin{align}
\rho=\frac{m-x}{m^3-m}\mathds{1}+\frac{mx-1}{m^3-3}  \mathcal{S}; ~~~~x\in [-1,1].
\end{align}
where $\mathcal{S}=\sum_{\mu, \nu} |\mu\rangle  \langle \nu| \otimes |\nu\rangle  \langle \mu| $ is an exchange operator. 
\begin{figure*}[!ht]
\centering\includegraphics[width=0.5\linewidth]{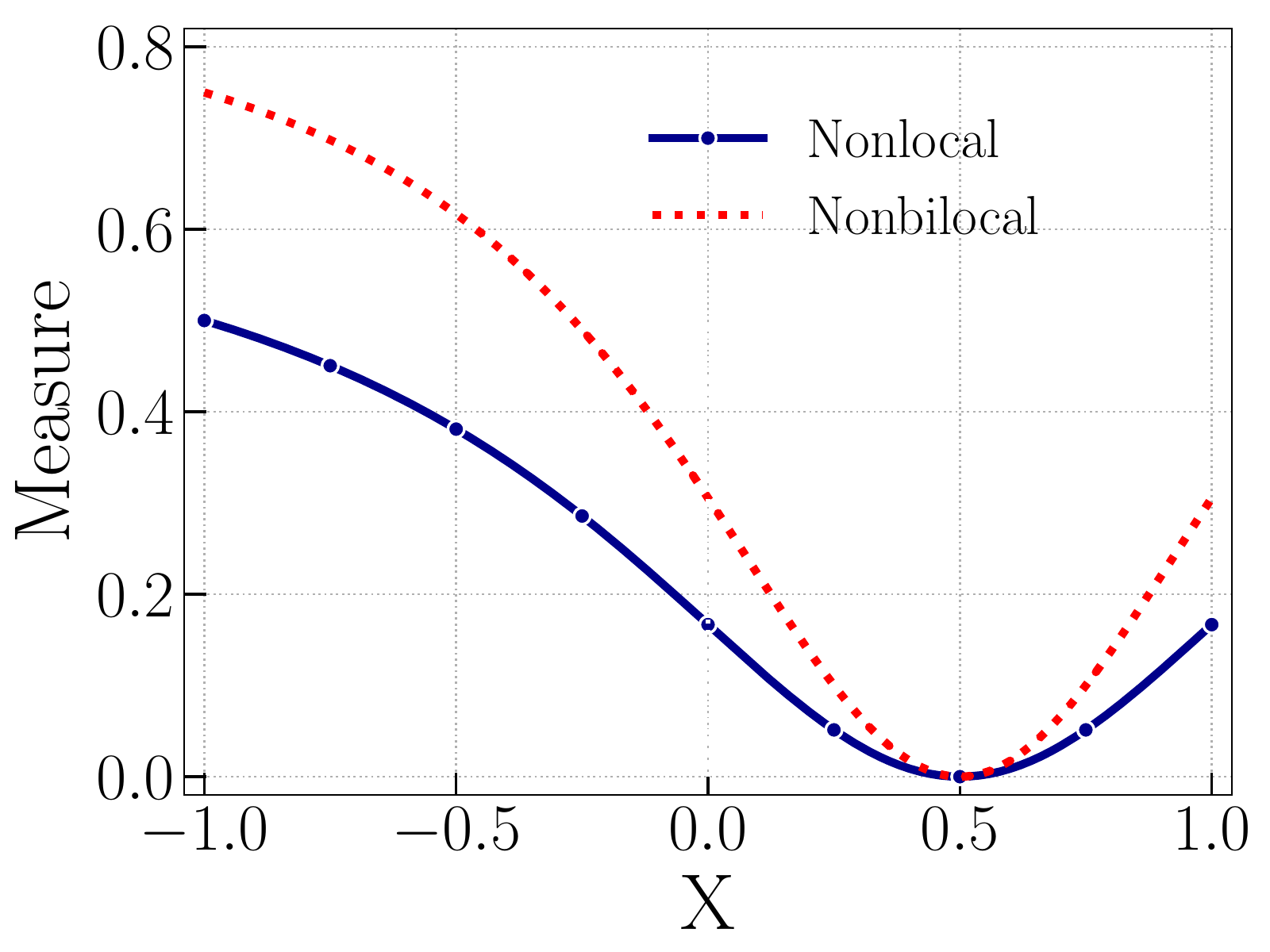}
\caption{(color online) The fidelity based nonlocal and nonbilocal measure for $2\times 2$ isotropic state. }
\label{fig1}
\end{figure*}
We observe that the F-MIN of Werner state vanishes at  $x=1/m$. In Fig. (2), we have plotted the F-MIN and fidelity based nonbilocal measure as a function of state parameter $x$. We clearly notice that the nonbilocal measure is also zero at $x=1/m$ and is always greater than the F-MIN.

\section{Conclusions}
\label{Concl}
In this article, we have introduced a measure of nonbilocal correlations of two bipartite input states using fidelity based measurement-induced nonlocality. We notice that a closer connection exists between the nonlocal and nonbilocal correlation measures. For any arbitrary pure input state, we have evaluated the nonbilocal correlation analytically. The upper bounds of the fidelity-based nonbilocal measure are also obtained for mixed input states. As an illustration, the nonbilocality is computed for some well-known examples. 

\section*{Acknowledgements}
SB wishes to thank the Council of Scientific and Industrial Research, Government of India for the financial support under Grant No. 03(1456)/19/EMR-II. RR wishes to thank DAE-NBHM and  (CSIR), Government of India  for the ~~financial supports under the schemes 02011/3/20/2020-R\&D-II and 03(1456)/19/EMR-II respectively.

%
%


\begin{thebibliography}{30}
\bibitem{Nielsen2010}
Nielsen, M., Chuang, I.: Quantum Computation and Quantum Information, Cam-bridge University Press, Cambridge (2010).

\bibitem{Girolami2014}
Girolami, D.: Observable measure of quantum coherence in finite dimensional systems. Phys. Rev. Lett. {\bf 113},  170401 (2014).

\bibitem{Baumgratz2014}
Baumgratz, T.,  Cramer, M.,  Plenio, M.B.: Quantifying coherence. Phys. Rev. Lett. {\bf 113},  140401 (2014).

\bibitem{Einstein}
Einstein, A., Podolsky, B., Rosen, N.: Can quantum-mechanical description of physical reality be
considered complete?. Phys. Rev. {\bf 47}, 777 (1935).

\bibitem{scho}
Schrodinger, E.: Discussion of probability relations between separated systems.  Proc. Cambridge Philos. Soc. {\bf 31}, 555 (1935).
	
\bibitem{Bell} 
Bell, J.S.: On the Einstein Podolsky Rosen paradox. Physics {\bf 1}, 195 (1964).
 
\bibitem{Ollivier2001}
Ollivier, H.,  Zurek, W.H.: Quantum discord: a measure of the quantumness of correlations. Phys. Rev. Lett. {\bf 88},  017901 (2001).
	
\bibitem{Dakic}
Dakic, B.,  Vedral, V., Brukner, C.: Necessary and sufficient condition for nonzero quantum discord. Phys. Rev. Lett. {\bf 105}, 190502 (2010).
 
\bibitem{Luo2010pra}
Luo, S., Fu, S.: Geometric measure of quantum discord. Phys. Rev. A {\bf 82},  034302 (2010).	
	
\bibitem{Luo2011}
Luo, S., Fu, S.: Measurement-induced nonlocality. Phys. Rev. Lett. {\bf 106}, 120401 (2011).
	
\bibitem{Buscemi}
Buscemi, F.: All entangled quantum states are nonlocal. Phys. Rev. Lett. {\bf 108}, 200401 (2012).

\bibitem{Werner}
 Werner, R. F.: Quantum states with Einstein-Podolsky-Rosen correlations admitting a hidden-variable model. Phys. Rev. A {\bf 40}, 4277 (1989).

\bibitem{Almeida}
Almeida,  M.  L.,  Pironio, S.,  Barrett, J., Toth, G., Acin, A.: Noise robustness of the nonlocality of entangled quantum states. Phys. Rev. Lett. {\bf 99}, 040403 (2007).

\bibitem{Branciard2010}
Branciard, C., Gisin, N., Pironio, S.: Characterizing the nonlocal correlations of particles that never interacted. Phys. Rev. Lett. {\bf 104}, 170401 (2010).

\bibitem{Branciard2012}
Branciard, C., Rosset, D., Gisin, N., Pironio, S.: Bilocal versus nonbilocal correlations in entanglement-swapping experiments.  Phys. Rev. A {\bf 85}, 032119 (2012).

\bibitem{Fritz2012}
Fritz, T.: Beyond Bell's theorem: correlation scenarios. New J. Phys. {\bf 14}, 103001 (2012).

\bibitem{Fritz2016}
Fritz, T.: Beyond Bell's Theorem II: Scenarios with arbitrary causal structure.  Comm. Math. Phys. {\bf 341}, 391-434 (2016).

\bibitem{Wood2015}
Wood, C. J.,  Spekkens, R. W.: The lesson of causal discovery algorithms for quantum correlations: causal explanations of Bell-inequality violations require fine-tuning. New J. Phys. {\bf 17}, 033002 (2015).

\bibitem{Henson}
Henson, J., Lal, R.,  Pusey, M. F.: Theory-independent limits on correlations from generalized Bayesian networks. New J. Phys. {\bf 16}, 113043 (2014).

\bibitem{Chaves2015}
Chaves,R.,  Brask, J. B., Brunner, N.: Device-Independent Tests of Entropy. Phys. Rev. Lett. {\bf 115}, 110501 (2015).

\bibitem{Tavakoli2014}
Tavakoli, A.,  Skrzypczyk, P.,  Cavalcanti, D.,  A. Acín, A.: Nonlocal correlations in the star-network configuration. Phys. Rev. A {\bf 90}, 062109 (2014).

\bibitem{Tavakoli2016a}
Tavakoli, A.: Quantum Correlations in connected multipartite Bell experiments. J. Phys. A: Math. Theor. {\bf 49}, 145304 (2016).

\bibitem{Tavakoli2016b}
Tavakoli, A.: Bell-type inequalities for arbitrary noncyclic networks.  Phys. Rev. {\bf A 93}, 030101 (2016).

\bibitem{Rosset}
Rosset, D., Branciard, C.,  Barnea, T. J., Putz, G., Brunner, N., Gisin, N.: Nonlinear Bell inequalities tailored for quantum networks. Phys. Rev. Lett. {\bf 116}, 010403 (2016).

\bibitem{Gisin2017}
Gisin, N., Mei, Q. X.,  Tavakoli, A., Renou, M. O., Brunner, N.: All entangled pure quantum states violate the bilocality inequality.  Phys. Rev. A {\bf 96}, 020304 (2017).

\bibitem{Palazuelos}
Palazuelos, C.: Super activation of quantum nonlocality. Phys. Rev. Lett. {\bf 109}, 190401 (2012).

\bibitem{Cavalcanti2011}
Cavalcanti, D.,   Almeida, M. L.,  Scarani,  V.,  Acin, A.: Quantum networks reveal quantum nonlocality. Nature Commun. {\bf 2}, 184 (2011).

\bibitem{Cavalcanti2012}
Cavalcanti, D.,  Rabelo, R., Scarani, V.: Nonlocality tests enhanced by a third observer. Phys. Rev. Lett. {\bf 108}, 040402 (2012).

\bibitem{Masanes}
 Masanes, L.,  Liang, Y.-C.,  Doherty, A. C.: All bipartite entangled states display some hidden nonlocality. Phys. Rev.Lett. {\bf 100}, 090403 (2008).

\bibitem{Lee2003}
Lee, J., Kim, M. S., Brukner, C.: Operationally invariant measure of the distance between quantum states by complementary measurements. Phys. Rev. Lett. {\bf  91}, 087902 (2003).

\bibitem{Cincio2018}
Cincio, L., Suba, Y.,  Sornborger, A. T.,  Coles, P. J.: Learning the quantum algorithm for state overlap. New J. Phys. {\bf 20}, 113022 (2018).

\bibitem{Travnicek2019}
Travnicek, V.,  Bartkiewicz, K., Cernoch, A., Lemr, K.: Experimental measurement of Hilbert-Schmidt distance between two-qubit states as means for speeding-up machine learning. Phys. Rev. Lett. {\bf 123}, 260501 (2019).

\bibitem{Pandya2000}
Pandya, P., Sakarya, O.,  Wie, M.: Hilbert-Schmidt distance and entanglement witnessing. Phys. Rev. A {\bf 102}, 012409 (2020).

\bibitem{Indrajith2021}
Indrajith,  V. S., Muthuganesan, R., Sankaranarayanan, R.: Measurement-induced nonlocality quantified by Hellinger distance and weak measurements. Physica A {\bf 566}, 125615 (2021).

\bibitem{Yao2016}
Yao, Y., Dong, G. H., Xiao, X., Sun, C.P.: Frobenius-norm-based measures of quantum coherence and asymmetry. Sci. Rep. {\bf 6}, 32010 (2016).

\bibitem{Dodonov2000}
Dodonov, V. V., Manko, O. V.,  Manko, V. I.,  Wünsche, A.: Hilbert-Schmidt distance and non-classicality of states in quantum optics.  J. Mod. Opt. {\bf 47}, 633 (2000).

\bibitem{Piani2012}
Piani, M.: Problem with geometric discord. Phys. Rev. A {\bf 86}, 034101 (2012).

\bibitem{Hu2015}
 Hu, M.-L., Fan, H.: Measurement-induced nonlocality based on the trace norm. New J. Phys. {\bf 17},  033004 (2015).

\bibitem{Jozsa1994}
Jozsa, R.: Fidelity for mixed states. J. Mod. Opt. {\bf 41},  2315 (1994).

\bibitem{Gisin1997}
Gisin, N.,  Massar, S.: Optimal quantum cloning machines. Phys. Rev. Lett. {\bf 79},  2153 (1997).

\bibitem{Zhang2007}
Zhang, G.-F.: Thermal entanglement and teleportation in a two-qubit Heisenberg chain with Dzyaloshinski-Moriya anisotropic antisymmetric interaction. Phys. Rev. A {\bf 75},  034304 (2007).

\bibitem{Gorin2006}
 Gorin, T.,  Prosen, T.,  Seligman, H., Znidaric, M.: Dynamics of Loschmidt echoes and fidelity decay. Phys. Rep. {\bf 435},  33 (2006).

\bibitem{Gu2010}
 Gu, S.-J.: Fidelity approach to quantum phase transitions. Int. J. Mod. Phys. B {\bf 24},  4371 (2010).

\bibitem{Wang}
 Wang, X. Yu, C.-S.,  Yi,  X.X.: An alternative quantum fidelity for mixed states of qudits. Phys. Lett. A {\bf 373}, 58 (2008).

\bibitem{MuthuPLA}
Muthuganesan, R, Sankaranarayanan, R.: Fidelity based measurement induced nonlocality and its dynamics in quantum noisy channels. Phys. Lett. A {\bf 381},  3855 (2017).

 \bibitem{MuthuPLA2}
Muthuganesan, R, Sankaranarayanan, R.: Fidelity based measurement induced nonlocality.  Phys. Lett. A {\bf 381},  3028 (2017).


\end{thebibliography}


\end{document}